\documentclass[12pt]{article}
\usepackage{amsmath,amssymb,amsthm,amsxtra,overpic,bbm,bm,epsfig,ulem}
\usepackage{color}
\usepackage{cite}

\usepackage{graphicx}

\usepackage{hyperref}
\usepackage{url}
\usepackage{multirow}
\usepackage{textcomp}

\def\thefootnote{\fnsymbol{footnote}}

\begin{document}

\vspace{0.2cm}

\begin{center}
{\large\bf Leptonic first-row correlation and unitarity waiting for further JUNO tests}
\end{center}

\vspace{0.2cm}

\begin{center}
{\bf Zhi-zhong Xing$^{1,2,3}$}
\footnote{E-mail: xingzz@ihep.ac.cn}
\\
{\small
$^{1}$Institute of High Energy Physics, Chinese Academy of Sciences, Beijing 100049, China \\
$^{2}$School of Physical Sciences,
University of Chinese Academy of Sciences, Beijing 100049, China \\
$^{3}$Center of High Energy Physics, Peking University, Beijing 100871, China}
\end{center}

\vspace{0.4cm}

\begin{abstract}
We conjecture that there exists a remarkable correlation among the three
elements in the first row of the $3\times 3$ lepton flavor mixing matrix $U$:
$|U^{}_{e1}|^2 = 2 \left(|U^{}_{e2}|^2 + |U^{}_{e3}|^2\right)$, which holds
even though $U$ is non-unitary in the canonical seesaw mechanism.
This ``first-row correlation" is fully consistent with
$\sin^2\theta^{}_{12} = \left(1 - 2\tan^2\theta^{}_{13}\right)/3$ in the
unitarity limit of $U$, as supported by the latest JUNO and Daya
Bay precision measurements at a confidence level close to $1\sigma$. 
\end{abstract}

\vspace{0.7cm}

\def\thefootnote{\arabic{footnote}}
\setcounter{footnote}{0}


\newpage 

As the world's largest detector for studying reactor antineutrino oscillations,
the Jiangmen Underground Neutrino Observatory (JUNO) has recently reported
an unprecedented precision measurement of the lepton flavor mixing angle
$\theta^{}_{12}$ that dominates solar neutrinos oscillations:
$\sin^2\theta^{}_{12} = 0.3092 \pm 0.0087$ at the $1\sigma$ confidence
level~\cite{Abusleme:2025wem}. This robust result, together with the
world's best measurement of the smallest lepton flavor mixing angle
$\theta^{}_{13}$ in the Daya Bay reactor antineutrino oscillation experiment
(i.e., $\sin^2 2\theta^{}_{13} = 0.0851 \pm 0.0024$ at the $1\sigma$
level~\cite{DayaBay:2022orm}), indicates that the elements $U^{}_{e 1}$,
$U^{}_{e 2}$ and $U^{}_{e 3}$ in the
first row of the $3\times 3$ lepton flavor mixing matrix $U$ have all
been determined to an excellent degree of accuracy. However, such a
conclusion relies on the assumption that $U$ itself is unitary and thus
can be fully parameterized in terms of three Euler rotation angles
($\theta^{}_{12}$, $\theta^{}_{13}$, $\theta^{}_{23}$) and one or three
CP-violating phases (i.e., $\delta^{}_\nu$ for Dirac neutrinos or
$\delta^{}_\nu$, $\rho$, $\sigma$ for Majorana
neutrinos)~\cite{ParticleDataGroup:2024cfk}.

Different from the $3\times 3$ quark flavor mixing matrix, whose unitarity
is a direct consequence of the standard model (SM) of particle physics,
whether $U$ is unitary or not depends on how the three
active neutrinos $\nu^{}_\alpha$ (for $\alpha = e, \mu, \tau$) acquire their
tiny masses beyond the SM. Given the widely recognized canonical seesaw
mechanism~\cite{Minkowski:1977sc} as the most natural and economical
extension of the SM to generate tiny Majorana masses for $\nu^{}_\alpha$,
$U$ must be non-unitary to some extent. The reason is that the seesaw
mechanism demands nonzero active-sterile flavor mixing between light and
heavy Majorana neutrinos which results from their Yukawa interactions with
the SM Higgs field, and hence gives rise to an unavoidable departure of
$U$ from its unitarity limit. In particular, this picture is fully consistent
with the spirit of Weinberg's SM effective field theories~\cite{Weinberg:1979sa}
after the heavy seesaw degrees of freedom are integrated out. A recent global
analysis of various currently available electroweak precision data and
neutrino oscillation data {\it indirectly} constrains the overall
non-unitarity of $U$ to be below ${\cal O}(10^{-2})$ in the seesaw
framework~\cite{Blennow:2023mqx}. That is why $U$
has simply been assumed to be unitary as a good leading-order approximation
in most of today's studies of massive neutrinos and their phenomenology.

To welcome the arrival of the precision measurement era of neutrino physics,
we are well motivated to ask whether and to what extent the JUNO experiment
can {\it directly} test the first-row unitarity of $U$ and any possible
correlation among $|U^{}_{e i}|$ (for $i = 1, 2, 3$) predicted by some viable
flavor symmetry models. This {\it Short Communication} is intended to answer
these two important questions. First, we are going to set a remarkable lower
bound on $|U^{}_{e1}|^2 + |U^{}_{e2}|^2 + |U^{}_{e3}|^2$ with the help of
Ref.~\cite{Blennow:2023mqx} in a complete Euler-like parametrization of the
whole seesaw flavor structure, so as to give an {\it indirect} benchmark for
the future {\it direct} constraints on this sum. We point out that it seems
unlikely for the JUNO experiment to constrain the first-row non-unitarity of
$U$ in the seesaw framework as the $\overline{\nu}^{}_e \to \overline{\nu}^{}_e$
oscillation probability is insensitive to this effect,
but the JUNO precision measurement is able to extract the Euler
angles $\theta^{}_{12}$ and $\theta^{}_{13}$ of $U$ that are uncontaminated
by nonzero active-sterile flavor mixing. Second, we shall
show that the present JUNO and Daya Bay data support a novel relation
$|U^{}_{e1}|^2 = 2 \left(|U^{}_{e2}|^2 + |U^{}_{e3}|^2\right)$ at a confidence
level close to $1\sigma$, even though $U$ itself is not exactly unitary. It
is expected that the forthcoming JUNO precision data will test this inspiring
first-row correlation to a much better degree of accuracy.


Without loss of generality, let us assume the existence of three right-handed
neutrino fields $N^{}_{\alpha \rm R}$ which mix with the SM left-handed
neutrino fields $\nu^{}_{\alpha \rm L}$ (for $\alpha = e, \mu, \tau$)
via the Yukawa interactions in the seesaw scheme~\cite{Minkowski:1977sc}.
The mass eigenstates of these active and sterile Majorana neutrinos are
denoted as $\nu^{}_i$ and $N^{}_i$ (for $i = 1, 2, 3$),
respectively. In the mass basis, the weak charged-current interactions
of three charged leptons and six massive neutrinos can be written as
\footnote{Note that $U$ and $R$ are the two $3 \times 3$ submatrices in the first
row of a $6 \times 6$ unitary matrix ${\cal U}$ used to diagonalize
the $6 \times 6$ overall mass matrix of active and sterile neutrinos in
the basis where the flavor eigenstates of three charged leptons are
identified with their mass eigenstates~\cite{Xing:2007zj,Xing:2011ur}.
They are therefore interwined with the other two $3 \times 3$ submatrices
in the second row of ${\cal U}$ as demanded by the normalization and
orthogonality conditions of ${\cal U}$ itself. But these two ``hidden"
matrices do not affect the relevant physics under discussion, simply because
they have nothing to do with ${\cal L}^{}_{\rm cc}$ in Eq.~(\ref{1}).}
\begin{eqnarray}
-{\cal L}^{}_{\rm cc} = \frac{g}{\sqrt{2}} \hspace{0.1cm}
\overline{\big(\begin{matrix} e & \mu & \tau\end{matrix}\big)^{}_{\rm L}}
\hspace{0.1cm} \gamma^\mu \left[ U \left( \begin{matrix} \nu^{}_{1}
\cr \nu^{}_{2} \cr \nu^{}_{3} \cr\end{matrix}
\right)^{}_{\hspace{-0.08cm} \rm L}
+ R \left(\begin{matrix} N^{}_1 \cr N^{}_2 \cr N^{}_3
\cr\end{matrix}\right)^{}_{\hspace{-0.08cm} \rm L} \hspace{0.05cm} \right]
W^-_\mu + {\rm h.c.} \; ,
\label{1}
\end{eqnarray}
where $g$ is the weak coupling constant of the SM, the lepton flavor mixing
matrix $U = A U^{}_0$ and its sterile analog $R$ are correlated with each
other through both the overall seesaw unitarity condition
$U U^\dagger + R R^\dagger = A A^\dagger + R R^\dagger = I$ and the exact
seesaw relation $U D^{}_\nu U^T = -R D^{}_N R^T$ with
$D^{}_\nu = {\rm Diag}\{m^{}_1, m^{}_2, m^{}_3\}$ and
$D^{}_N = {\rm Diag}\{M^{}_1, M^{}_2, M^{}_3\}$ being the diagonal light
and heavy neutrino mass matrices~\cite{Xing:2007zj}. A full Euler-like
parametrization of the seesaw flavor structure gives~\cite{Xing:2011ur}:
\begin{eqnarray}
A \hspace{-0.2cm} & = & \hspace{-0.2cm}
\left( \begin{matrix} c^{}_{14} c^{}_{15} c^{}_{16} & 0 & 0
\cr \vspace{-0.45cm} \cr
\begin{array}{l} -c^{}_{14} c^{}_{15} \hat{s}^{}_{16} \hat{s}^*_{26} -
c^{}_{14} \hat{s}^{}_{15} \hat{s}^*_{25} c^{}_{26} \\
-\hat{s}^{}_{14} \hat{s}^*_{24} c^{}_{25} c^{}_{26} \end{array} &
c^{}_{24} c^{}_{25} c^{}_{26} & 0 \cr \vspace{-0.45cm} \cr
\begin{array}{l} -c^{}_{14} c^{}_{15} \hat{s}^{}_{16} c^{}_{26} \hat{s}^*_{36}
+ c^{}_{14} \hat{s}^{}_{15} \hat{s}^*_{25} \hat{s}^{}_{26} \hat{s}^*_{36} \\
- c^{}_{14} \hat{s}^{}_{15} c^{}_{25} \hat{s}^*_{35} c^{}_{36} +
\hat{s}^{}_{14} \hat{s}^*_{24} c^{}_{25} \hat{s}^{}_{26}
\hat{s}^*_{36} \\
+ \hat{s}^{}_{14} \hat{s}^*_{24} \hat{s}^{}_{25} \hat{s}^*_{35}
c^{}_{36} - \hat{s}^{}_{14} c^{}_{24} \hat{s}^*_{34} c^{}_{35}
c^{}_{36} \end{array} &
\begin{array}{l} -c^{}_{24} c^{}_{25} \hat{s}^{}_{26} \hat{s}^*_{36} -
c^{}_{24} \hat{s}^{}_{25} \hat{s}^*_{35} c^{}_{36} \\
-\hat{s}^{}_{24} \hat{s}^*_{34} c^{}_{35} c^{}_{36} \end{array} &
c^{}_{34} c^{}_{35} c^{}_{36} \cr \end{matrix} \right) \; , \hspace{0.5cm}
\nonumber \\
R \hspace{-0.2cm} & = & \hspace{-0.2cm}
\left( \begin{matrix} \hat{s}^*_{14} c^{}_{15} c^{}_{16} &
\hat{s}^*_{15} c^{}_{16} & \hat{s}^*_{16} \cr \vspace{-0.45cm} \cr
\begin{array}{l} -\hat{s}^*_{14} c^{}_{15} \hat{s}^{}_{16} \hat{s}^*_{26} -
\hat{s}^*_{14} \hat{s}^{}_{15} \hat{s}^*_{25} c^{}_{26} \\
+ c^{}_{14} \hat{s}^*_{24} c^{}_{25} c^{}_{26} \end{array} & -
\hat{s}^*_{15} \hat{s}^{}_{16} \hat{s}^*_{26} + c^{}_{15}
\hat{s}^*_{25} c^{}_{26} & c^{}_{16} \hat{s}^*_{26} \cr \vspace{-0.45cm} \cr
\begin{array}{l} -\hat{s}^*_{14} c^{}_{15} \hat{s}^{}_{16} c^{}_{26}
\hat{s}^*_{36} + \hat{s}^*_{14} \hat{s}^{}_{15} \hat{s}^*_{25}
\hat{s}^{}_{26} \hat{s}^*_{36} \\ - \hat{s}^*_{14} \hat{s}^{}_{15}
c^{}_{25} \hat{s}^*_{35} c^{}_{36} - c^{}_{14} \hat{s}^*_{24}
c^{}_{25} \hat{s}^{}_{26}
\hat{s}^*_{36} \\
- c^{}_{14} \hat{s}^*_{24} \hat{s}^{}_{25} \hat{s}^*_{35}
c^{}_{36} + c^{}_{14} c^{}_{24} \hat{s}^*_{34} c^{}_{35} c^{}_{36}
\end{array} &
\begin{array}{l} -\hat{s}^*_{15} \hat{s}^{}_{16} c^{}_{26} \hat{s}^*_{36}
- c^{}_{15} \hat{s}^*_{25} \hat{s}^{}_{26} \hat{s}^*_{36} \\
+c^{}_{15} c^{}_{25} \hat{s}^*_{35} c^{}_{36} \end{array} &
c^{}_{16} c^{}_{26} \hat{s}^*_{36} \cr \end{matrix} \right) \; ,
\hspace{0.5cm}
\label{2}
\end{eqnarray}
where $c^{}_{ij} \equiv \cos\theta^{}_{ij}$, $s^{}_{ij} \equiv \sin\theta^{}_{ij}$
and $\hat{s}^{}_{ij} \equiv s^{}_{ij} e^{{\rm i}\delta^{}_{ij}}$ with
$\theta^{}_{ij}$ and $\delta^{}_{ij}$ being the active-sterile flavor mixing
angles and CP-violating phases (for $i = 1, 2, 3$ and $j = 4, 5, 6$), and
\begin{eqnarray}
U^{}_0 = \left( \begin{matrix}
e^{-{\rm i}\rho} & 0 & 0 \cr
0 & e^{-{\rm i}\sigma} & 0 \cr
0 & 0 & 1 \cr\end{matrix} \right)
\left( \begin{matrix} c^{}_{12} c^{}_{13} & s^{}_{12}
c^{}_{13} & s^{}_{13} e^{-{\rm i}\delta^{}_\nu} \cr
-s^{}_{12} c^{}_{23} - c^{}_{12} s^{}_{13} s^{}_{23} e^{{\rm i}\delta^{}_\nu}
& c^{}_{12} c^{}_{23} -
s^{}_{12} s^{}_{13} s^{}_{23} e^{{\rm i}\delta^{}_\nu} & c^{}_{13} s^{}_{23} \cr
s^{}_{12} s^{}_{23} - c^{}_{12} s^{}_{13} c^{}_{23} e^{{\rm i}\delta^{}_\nu}
& -c^{}_{12} s^{}_{23} - s^{}_{12} s^{}_{13} c^{}_{23} e^{{\rm i}\delta^{}_\nu}
& c^{}_{13} c^{}_{23}
\cr \end{matrix} \right)
\left( \begin{matrix}
e^{{\rm i}\rho} & 0 & 0 \cr
0 & e^{{\rm i}\sigma} & 0 \cr
0 & 0 & 1 \cr\end{matrix} \right) \;
\label{3}
\end{eqnarray}
is the unitary part of $U$ which
represents a leptonic analog of the $3\times 3$ quark flavor mixing matrix but
consists of three CP-violating phases (i.e., $\delta^{}_\nu \equiv \delta^{}_{13}
- \delta^{}_{12} - \delta^{}_{23}$, $\rho \equiv \delta^{}_{12} + \delta^{}_{23}$
and $\sigma \equiv \delta^{}_{23}$). It is then straightforward to see that the
three elements in the first row of $U = A U^{}_0$ read as
\begin{eqnarray}
\big\{U^{}_{e1} \; , \hspace{0.25cm} U^{}_{e2} \; ,
\hspace{0.25cm} U^{}_{e3}\big\}
= c^{}_{14} c^{}_{15} c^{}_{16} \times \big\{ c^{}_{12} c^{}_{13} \; ,
\hspace{0.25cm} \hat{s}^*_{12} c^{}_{13} \; ,
\hspace{0.25cm} \hat{s}^*_{13} \big\} \; ,
\label{4}
\end{eqnarray}
implying that its first-row non-unitarity in the seesaw mechanism is
simply characterized by the active-sterile flavor mixing angles
$\theta^{}_{1j}$ (for $j = 4, 5, 6$) as follows:
\begin{eqnarray}
\sum^3_{i=1} \left|U^{}_{e i}\right|^2
= 1 - \sum^3_{i=1} |R^{}_{e i}|^2 = \prod^6_{j=4}
c^2_{1j} = 1 - \sum^6_{j=4} s^2_{1j} + \cdots \; ,
\label{5}
\end{eqnarray}
where ``$\cdots$" denotes the higher-order terms of $s^{}_{1j}$.
So a {\it direct} precision measurement of the sum on the left-hand
side of Eq.~(\ref{5}) will constrain to what extent $\theta^{}_{14}$,
$\theta^{}_{15}$ and $\theta^{}_{16}$ may deviate from zero. On the
other hand, the constraints on these three original seesaw flavor
parameters will allow one to derive an {\it indirect} lower bound
on the first-row non-unitarity of $U$.

In the seesaw framework under discussion, some strong limits on the
deviation of $U$ from $U^{}_0$ (i.e., the deviation of $A$ from $I$)
have been inferred from a recent global analysis of the electroweak
precision measurements and the precision data on quark flavor
physics and neutrino oscillations~\cite{Blennow:2023mqx}:
\begin{eqnarray}
& {\rm NMO:} & \left| I - A\right| < \left( \begin{matrix}
1.3 \times 10^{-3} & 0 & 0 \cr
2.4 \times 10^{-5} & 1.1 \times 10^{-5} & 0 \cr
1.8 \times 10^{-3} & 1.1 \times 10^{-4} & 1.0 \times 10^{-3} \cr \end{matrix}
\right) \; ,
\nonumber \\
& {\rm IMO:} & \left| I - A\right| < \left( \begin{matrix}
1.4 \times 10^{-3} & 0 & 0 \cr
2.4 \times 10^{-5} & 1.0 \times 10^{-5} & 0 \cr
1.6 \times 10^{-3} & 3.6 \times 10^{-5} & 8.1 \times 10^{-4} \cr \end{matrix}
\right) \; , \hspace{1.1cm}
\label{6}
\end{eqnarray}
at the $95\%$ confidence level for the normal mass ordering
(NMO, $m^{}_1 < m^{}_2 < m^{}_3$) or the inverted mass ordering (IMO,
$m^{}_3 < m^{}_1 < m^{}_2$) of three
light Majorana neutrinos. These constraints can be translated
into the upper bounds on the active-sterile flavor mixing angles
$\theta^{}_{1j}$, $\theta^{}_{2j}$ and $\theta^{}_{3j}$
(for $j = 4, 5, 6$)~\cite{Xing:2024gmy}:
$\theta^{}_{1j} < 2.92^\circ$,
$\theta^{}_{2j} < 0.27^\circ$ and
$\theta^{}_{3j} < 2.56^\circ$ (NMO); or
$\theta^{}_{1j} < 3.03^\circ$,
$\theta^{}_{2j} < 0.26^\circ$ and
$\theta^{}_{3j} < 2.31^\circ$ (IMO). So we infer the following {\it indirect}
constraints on the first-row non-unitarity of $U$ from Eq.~(\ref{5}):
\begin{eqnarray}
\left|U^{}_{e1}\right|^2 + \left|U^{}_{e2}\right|^2 + \left|U^{}_{e3}\right|^2
> \left\{
\begin{array}{l}
\hspace{-0.1cm} \displaystyle
0.9974 \; \quad \left({\rm for} ~ m^{}_1 < m^{}_2 < m^{}_3\right) \; , \\
\hspace{-0.1cm} \displaystyle
0.9972 \; \quad \left({\rm for} ~ m^{}_3 < m^{}_1 < m^{}_2\right) \; ,
\end{array} \right.
\label{7}
\end{eqnarray}
at the same confidence level. One may then conclude that the normalization
condition for the first row of $U$ has been tested to the sensitivity
of about $3 \times 10^{-3}$ --- a striking degree of accuracy comparable
with the robust result
$\left| V^{}_{ud}\right|^2 + \left| V^{}_{us}\right|^2 +
\left| V^{}_{ub}\right|^2 = 0.9984 \pm 0.0007$ obtained from a global test
of the first-row normalization condition of the
$3\times 3$ quark flavor mixing matrix~\cite{ParticleDataGroup:2024cfk}.

To what extent can the direct precision measurements of
$\left|U^{}_{e1}\right|^2 + \left|U^{}_{e2}\right|^2
+ \left|U^{}_{e3}\right|^2$ in the JUNO and Daya Bay experiments help to
test the first-row unitarity of $U$? To answer this question, let us take
a look at the probability of reactor antineutrino oscillations
${\cal P}\left(\overline{\nu}^{}_e \to \overline{\nu}^{}_e\right)$.
Without concern about the weak charged-current interactions associated
with both the source and detector, we find that
${\cal P}\left(\overline{\nu}^{}_e \to \overline{\nu}^{}_e\right)$ as a
function of the average beam energy $E$ and the baseline length $L$
can be expressed as
\footnote{Note that the heavy Majorana neutrinos are kinematically
forbidden to take part in the flavor oscillations of reactor antineutrinos.
A careful analysis shows that terrestrial matter effects in the JUNO
experiment is expected to be negligibly small~\cite{Li:2025hye}, and thus
will not be taken into account in this work.}
\begin{eqnarray}
{\cal P}\left(\overline{\nu}^{}_e \to \overline{\nu}^{}_e\right)
\hspace{-0.2cm} & = & \hspace{-0.2cm}
\frac{1}{\displaystyle\left[\left(U U^\dagger\right)_{ee}\right]^2}
\left[\left(\sum^3_{i=1} \left|U^{}_{ei}\right|^2\right)^2
- 4 \sum^3_{i<i^\prime} \left|U^{}_{ei}\right|^2
\left|U^{}_{ei^\prime}\right|^2 \sin^2\frac{\Delta m^2_{i^\prime i} L}
{4 E}\right] \hspace{0.5cm}
\nonumber \\
\hspace{-0.2cm} & = & \hspace{-0.2cm}
1 - 4 \left|\left(U^{}_0\right)^{}_{e1}\right|^2
\left|\left(U^{}_0\right)^{}_{e2}\right|^2
\sin^2 \frac{\Delta m^{2}_{21} L}{4 E}
\nonumber \\
\hspace{-0.2cm} & & \hspace{-0.2cm} \hspace{0.33cm}
- \hspace{0.05cm} 2 \left|\left(U^{}_0\right)^{}_{e3}\right|^2 \left(1 -
\left|\left(U^{}_0\right)^{}_{e3}\right|^2\right)
\left(\sin^2 \frac{\Delta m^{2}_{31} L}{4 E}
+ \sin^2 \frac{\Delta m^{2}_{32} L}{4 E}\right)
\nonumber \\
\hspace{-0.2cm} & & \hspace{-0.2cm} \hspace{0.33cm}
- \hspace{0.05cm} 2 \left|\left(U^{}_0\right)^{}_{e3}\right|^2
\left(\left|\left(U^{}_0\right)^{}_{e1}\right|^2
- \left|\left(U^{}_0\right)^{}_{e2}\right|^2 \right)
\sin \frac{\Delta m^{2}_{21} L}{4 E}
\sin \frac{\left(\Delta m^{2}_{31} + \Delta m^2_{32}\right) L}{4 E} \; ,
\hspace{0.5cm}
\label{8}
\end{eqnarray}
where the subscripts ``$i$" and ``$i^\prime$" run over $\left(1, 2, 3\right)$,
and $\Delta m^2_{i^\prime i} \equiv m^2_{i^\prime} - m^2_i$ denotes the neutrino
mass-squared difference. One can see that the non-unitarity effect hidden in the
first row of $U$ has been cancelled out in
${\cal P}\left(\overline{\nu}^{}_e \to \overline{\nu}^{}_e\right)$,
after Eqs.~(\ref{4}) and (\ref{5})
together with $\left(U U^\dagger\right)_{ee} = \left(A A^\dagger\right)_{ee} =
c^2_{14} c^2_{15} c^2_{16}$ are taken into account. Some comments are in order.
\begin{itemize}
\item     ${\cal P}\left(\overline{\nu}^{}_e \to \overline{\nu}^{}_e\right)$
is insensitive to the first-row non-unitarity of $U$, and its
$E$- and $L$-dependent terms in the second, third and fourth lines of
Eq.~(\ref{8}) correspond respectively to the leading
($\Delta m^2_{21}$-driven) JUNO oscillation behavior, the sub-leading
($\Delta m^2_{31}$- and $\Delta m^2_{32}$-driven) oscillations,
and the interference effect sensitive to the neutrino mass ordering
(i.e., to the sign of $\Delta m^2_{31} + \Delta m^2_{32}$).

\item     In reality, the JUNO reactor $\overline{\nu}^{}_e$ flux may involve
an overall uncertainty or the so-called zero-distance effect that can be
calibrated by the near detector TAO (Taishan Antineutrino Observatory)
at $L \to 0$~\cite{JUNO:2020ijm}. Then
the moduli of $\left(U^{}_0\right)^{}_{e1}$, $\left(U^{}_0\right)^{}_{e2}$
and $\left(U^{}_0\right)^{}_{e3}$ are all determinable from a precision
measurement of the $\overline{\nu}^{}_e$ spectrum at the JUNO far detector,
making it possible to test the unitarity condition
$|\left(U^{}_0\right)^{}_{e1}|^2 + |\left(U^{}_0\right)^{}_{e2}|^2
+ |\left(U^{}_0\right)^{}_{e3}|^2 = 1$ (similar to the test of
$\left| V^{}_{ud}\right|^2 + \left| V^{}_{us}\right|^2 +
\left| V^{}_{ub}\right|^2 = 1$).

\item     Even in the absence of a near detector, the JUNO experiment
is also able to cleanly measure $\theta^{}_{12}$ and $\theta^{}_{13}$ by
comparing between any two of the three oscillation amplitudes in Eq.~(\ref{8})
to cancel out the overall $\overline{\nu}^{}_e$ flux uncertainty. Namely,
$\left|\left(U^{}_0\right)^{}_{e1}\right|^2
\left|\left(U^{}_0\right)^{}_{e2}\right|^2
= \sin^2 2\theta^{}_{12} c^4_{13}/4$,
$\left|\left(U^{}_0\right)^{}_{e3}\right|^2 \left(1 -
\left|\left(U^{}_0\right)^{}_{e3}\right|^2\right)
= \sin^2 2\theta^{}_{13}/4$ and
$\left|\left(U^{}_0\right)^{}_{e3}\right|^2
\left(\left|\left(U^{}_0\right)^{}_{e1}\right|^2
- \left|\left(U^{}_0\right)^{}_{e2}\right|^2 \right)
= \sin^2 2\theta^{}_{13} \cos 2\theta^{}_{12}/4$. Note that determining the
third oscillation amplitude in Eq.~(\ref{8}) is the primary goal of the JUNO
detector, as this term is highly associated with a real measurement of the
neutrino mass ordering.
\end{itemize}
As the present JUNO and Daya Bay results for
$\theta^{}_{12}$~\cite{Abusleme:2025wem} and
$\theta^{}_{13}$~\cite{DayaBay:2022orm} are achieved in the assumption
of the $U \to U^{}_0$ (or $A \to I$) limit, they can be used to examine
whether there is a potential first-row correlation of $U^{}_0$
implied by an underlying flavor symmetry in the seesaw framework.

\begin{figure}[t]
\begin{center}
\includegraphics[width=8cm]{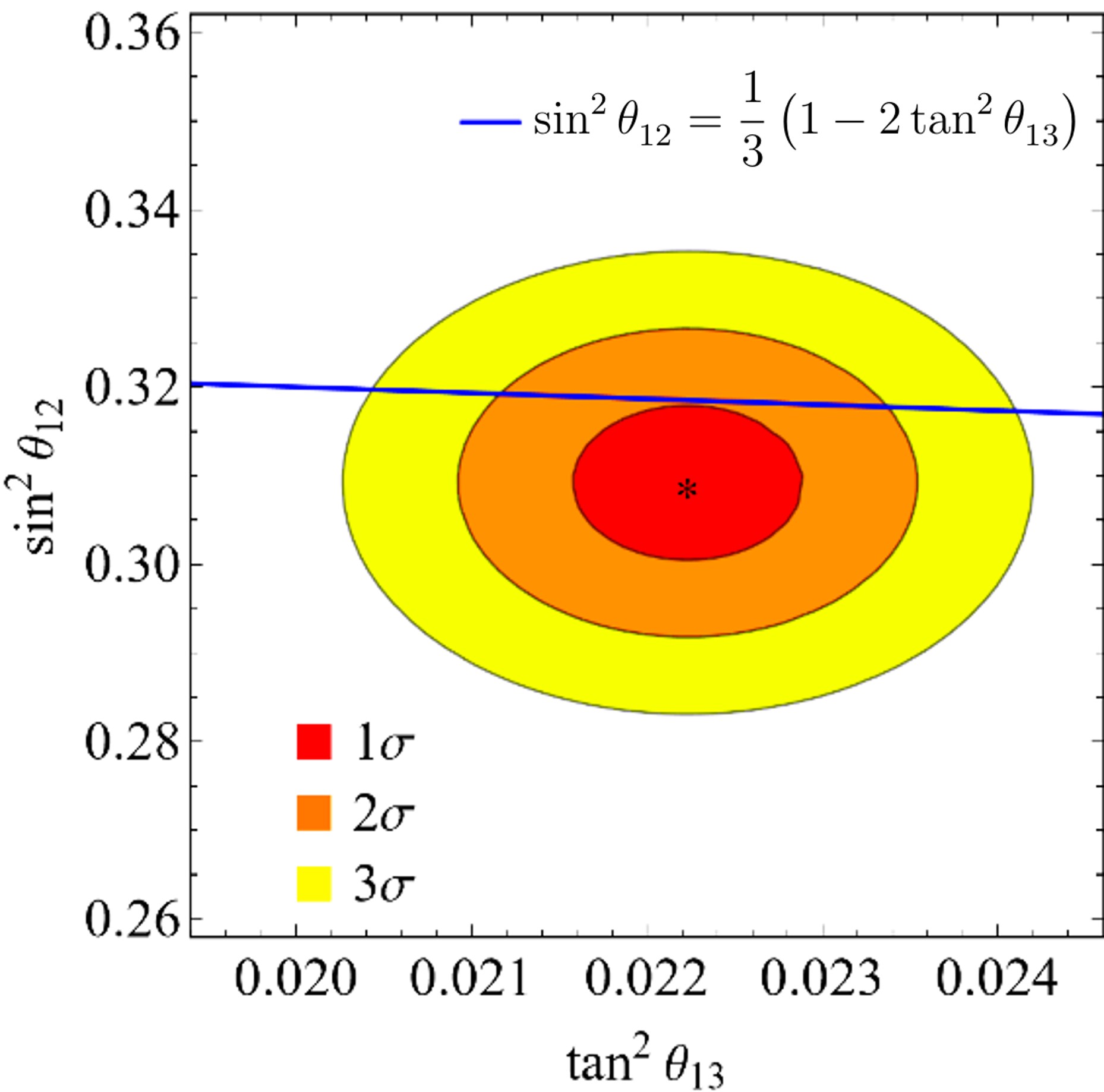}
\vspace{-0.25cm}
\caption{Confronting the JUNO~\cite{Abusleme:2025wem} and
Daya Bay~\cite{DayaBay:2022orm} precision data (the best-fit values and
the $1\sigma$, $2\sigma$ and $3\sigma$ regions) with a correlation
between $\theta^{}_{12}$ and $\theta^{}_{13}$ (the blue line) predicted
in Ref.~\cite{Xing:2006ms}.}
\label{Figure1}
\end{center}
\end{figure}

\begin{figure}[t]
\begin{center}
\includegraphics[width=9.6cm]{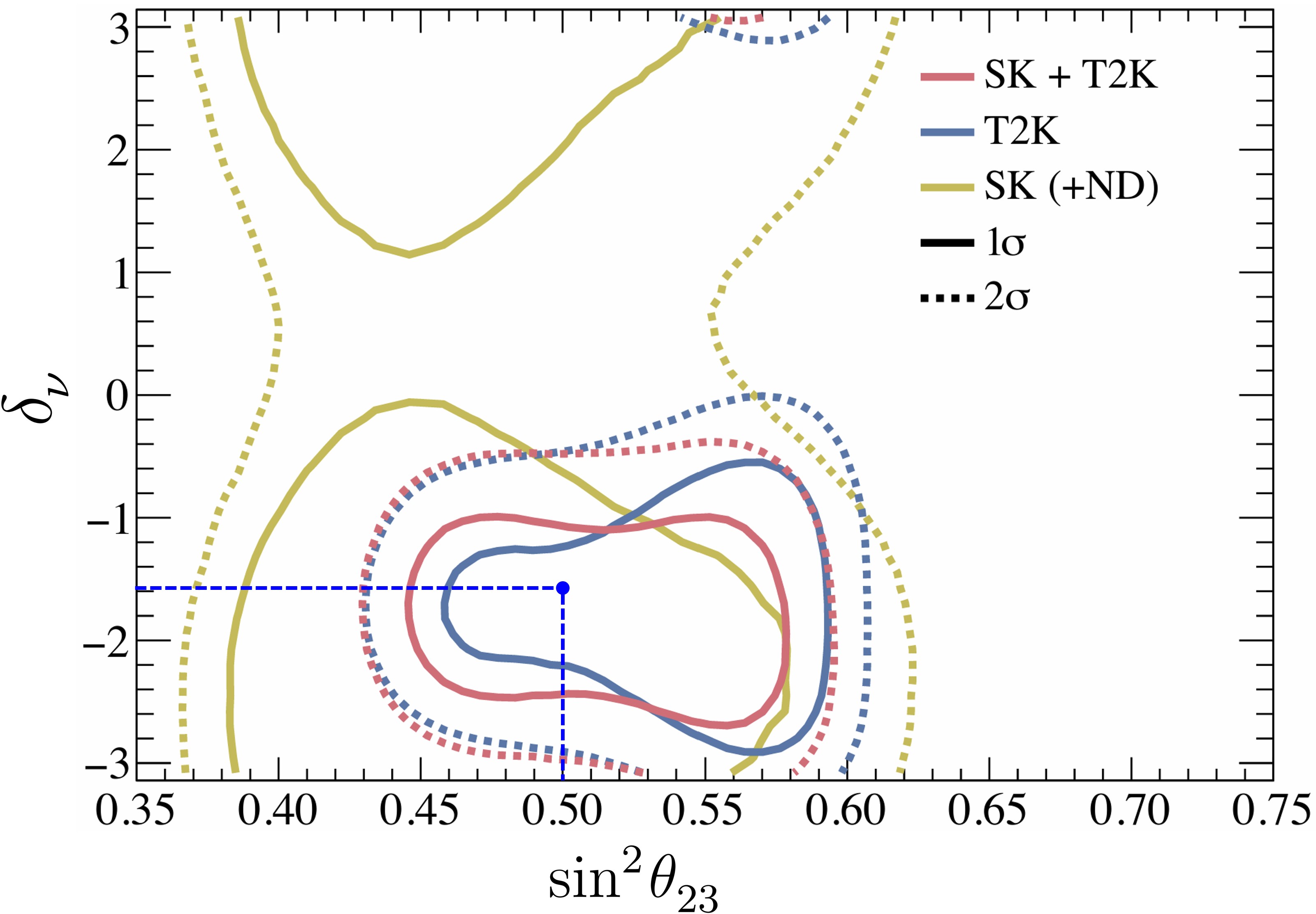}
\vspace{-0.25cm}
\caption{A recent joint analysis of SK (atmospheric) and T2K (accelerator)
neutrino oscillation data~\cite{T2K:2024wfn} favors $\theta^{}_{23} = \pi/4$
and $\delta^{}_\nu = -\pi/2$ (the blue point) for the standard parametrization
of $U^{}_0$.}
\label{Figure2}
\end{center}
\end{figure}

We find that the latest JUNO and Daya Bay precision data support the
following correlation between $\theta^{}_{12}$ and $\theta^{}_{13}$ at a
confidence level close to $1\sigma$:
\begin{eqnarray}
\sin^2\theta^{}_{12} = \frac{1}{3} \left(1 - 2\tan^2\theta^{}_{13}
\right) \; ,
\label{9}
\end{eqnarray}
as illustrated in Fig.~\ref{Figure1}. This remarkable correlation,
together with $\theta^{}_{23} = \pi/4$ and $\delta^{}_\nu = -\pi/2$, is
a natural consequence of the so-called TM1 lepton flavor mixing pattern of
$U^{}_0$ which was first proposed by Zhou and his collaborator in
Ref.~\cite{Xing:2006ms}. It is worth mentioning that the recent
T2K accelerator neutrino oscillation data, combined with the SK
atmospheric neutrino oscillation data, favor the possibilities of
$\theta^{}_{23} = \pi/4$ and $\delta^{}_\nu = -\pi/2$~\cite{T2K:2024wfn},
as can be seen in Fig.~\ref{Figure2}. That is why the TM1 pattern
of $U^{}_0$ has been regarded as one of the most likely leading forms of $U$
that can be naturally derived from some promising flavor symmetries in the
seesaw mechanism~\cite{Xing:2015fdg}
\footnote{We refer the reader to a few new examples of model building based on
the conventional or modular flavor symmetries in Refs.~\cite{Jiang:2025hvq,
Petcov:2025aci,Ding:2025dqd}, which emerged soon after the release of the JUNO
precision measurement of $\theta^{}_{12}$~\cite{Abusleme:2025wem}.}.

Note that Eq.~(\ref{9}) is simply equivalent to
$c^2_{12} c^2_{13} = 2\left(s^2_{12} c^2_{13} + s^2_{13}\right)$, and thus
equivalent to $\left|\left(U^{}_0\right)^{}_{e1}\right|^2 =
2 \left[\left|\left(U^{}_0\right)^{}_{e2}\right|^2
+ \left|\left(U^{}_0\right)^{}_{e3}\right|^2\right]$ for the unitary matrix
$U^{}_0$ as parameterized in Eq.~(\ref{3}). Such a relation, combined with the
first-row normalization condition of $U^{}_0$, leads us to
$\left|\left(U^{}_0\right)^{}_{e1}\right|^2 = 2/3$ as well.
Taking account of Eq.~(\ref{4}), we immediately find that this striking first-row
correlation of $U^{}_0$ can be generalized to an analogous first-row correlation
of $U$
at the same confidence level as Eq.~(\ref{9}) holds,
\begin{eqnarray}
\left|U^{}_{e1}\right|^2 = 2\left(\left|U^{}_{e2}\right|^2 +
\left|U^{}_{e3}\right|^2 \right) \; ,
\label{10}
\end{eqnarray}
even though $U$ itself is not exactly unitary. If the above phenomenological conjecture
can be further verified in the upcoming JUNO precision measurements, it will impose
a strong constraint on the texture of the active-sterile flavor mixing matrix $R$
via the exact seesaw relation $U D^{}_\nu U^T = -R D^{}_N R^T$. Thanks to such a
seesaw bridge, $R$ and $U$ are likely to share a similar flavor symmetry to assure the
consistency of model building in the leading order approximation.

Of course, any flavor symmetry behind a given pattern of $U$ or $U^{}_0$ must be
broken due to unavoidable quantum corrections from the seesaw scale down to the
electroweak scale.
As for the TM1 pattern of $U^{}_0$ discussed above, the flavor symmetry
breaking may bring $\theta^{}_{23}$ either to the upper octant ($\theta^{}_{23} > \pi/4$)
or to the lower one ($\theta^{}_{23} < \pi/4$), which is accordingly
favored in the NMO or IMO case as shown by a global analysis of all the currently
available neutrino oscillation data~\cite{Capozzi:2025wyn}. In view of
the very fact that the neutrino mass ordering will soon be probed in the JUNO experiment
as its primary target, we find it particularly interesting to examine whether there is
a potential correlation between the octant of $\theta^{}_{23}$ and the neutrino mass
ordering.
A careful numerical study along this line of thought has recently been done
with the help of the renormalization-group equations in the seesaw framework, indicating
that there is a possible parameter space to bring the TM1 flavor mixing pattern of
$U^{}_0$ to a better agreement with the existing JUNO, Daya Bay, SK and T2K
data~\cite{Zhang:2025jnn}.
This observation implies that the precision neutrino oscillation data will help a lot
to verify or disprove some typical lepton flavor symmetry models in the foreseeable future.

Although our discussions are subject to the first-row elements of $U$
and the reactor antineutrino oscillations, testing the correlation in Eq.~(\ref{10})
and constraining the non-unitarity of $U$ to much better degrees of accuracy call for
the synergy of various precision measurements of $|U^{}_{\alpha i}|$ (for
$\alpha = e, \mu, \tau$ and $i = 1, 2, 3$) to be implemented with the JUNO,
DUNE and Hyper-K detectors,
besides those existing neutrino oscillation experiments. In this case, of course, one
has to carefully take into account significant terrestrial matter effects
as they must be entangled with the non-unitarity effects associated with $U$ in the
accelerator-based DUNE and Hyper-K measurements, which are sensitive to all the three
neutrino flavors in both the appearance $\nu^{}_\mu \to \nu^{}_e$ and
$\nu^{}_\mu \to \nu^{}_\tau$ oscillation channels and the disappearance
$\nu^{}_\mu \to \nu^{}_\mu$ oscillation modes.

In conclusion, the JUNO precision measurement of $\theta^{}_{12}$ is a new milestone
of the reactor antineutrino oscillation experiments. Although the JUNO experiment
itself is quite unlikely to constrain the first-row non-unitarity of the $3\times 3$
lepton flavor mixing matrix $U$ in the seesaw framework
\footnote{The possibility of using elastic antineutrino-electron scattering to
probe the non-unitarity of $U$ in the JUNO and TAO experiments has recently been
studied~\cite{Huang:2025znh}, but this new approach is expected to be very challenging.},
it is able to extract the Euler angles $\theta^{}_{12}$ and $\theta^{}_{13}$ of $U$
that are uncontaminated by the active-sterile flavor mixing effects. The present
JUNO and Daya Bay data motivate us to conjecture an intriguing first-row correlation
$|U^{}_{e1}|^2 = 2 \left(|U^{}_{e2}|^2 + |U^{}_{e3}|^2\right)$. Such a phenomenological
conjecture and possible flavor symmetries behind it will be further tested in the era
of precision neutrino physics.

\vspace{0.5cm}

\textbf{Acknowledgments}

The author is greatly indebted to Yu-Feng Li and Shun Zhou for useful discussions,
and especially to Shun Zhou for his kind help in plotting Fig.~\ref{Figure1}. This
research was supported in part by the National Natural Science Foundation
of China (12535007), and  the Scientific and Technological
Innovation Program of the Institute of High Energy Physics (E55457U2).

\end{document}